\documentclass[groupedaddress,prx,amsmath,amssymb,aps,11pt]{revtex4-1}

\usepackage{graphicx}
\usepackage{dcolumn}
\usepackage{bm}
\usepackage{nicefrac}

\begin{document}

\title{Broken symmetries and Kohn's theorem in graphene cyclotron resonance}

\author{Jordan Pack} \altaffiliation[These authors contributed equally to this work.]{}
\author{B.\ Jordan Russell} \altaffiliation[These authors contributed equally to this work.]{}
\author{Yashika Kapoor}
\author{Jesse Balgley}
\author{Jeff Ahlers}
\affiliation{Department of Physics, Washington University in St.\ Louis, 1 Brookings Dr., St.\ Louis MO 63130, USA}
\author{Takashi Taniguchi}
\affiliation{International Center for Materials Nanoarchitectonics,
National Institute for Materials Science, 1-1 Namiki, Tsukuba 305-0044, Japan}
\author{Kenji Watanabe}
\affiliation{Research Center for Functional Materials,
National Institute for Materials Science, 1-1 Namiki, Tsukuba 305-0044, Japan}
\author{Erik A.\ Henriksen}
\email{henriksen@wustl.edu}
\affiliation{Department of Physics, Washington University in St.\ Louis, 1 Brookings Dr., St.\ Louis MO 63130, USA}
\affiliation{Institute for Materials Science \& Engineering, Washington University in St.\ Louis, 1 Brookings Dr., St.\ Louis MO 63130, USA}

\date{\today}

\begin{abstract}
The cyclotron resonance of monolayer graphene, encapsulated in hexagonal boron nitride and with a graphite back-gate, is explored via infrared transmission magnetospectroscopy as a function of the filling factor at fixed magnetic fields. The impact of many-particle interactions in the regime of broken spin and valley symmetries is observed spectroscopically.  As the occupancy of the zeroth Landau level is increased from half-filling, a non-monotonic progression of multiple cyclotron resonance peaks is seen for several interband transitions, and reveals the evolution of underlying many-particle-enhanced gaps. Analysis of the peak energies shows a significant exchange enhancements of spin gaps both at and below the Fermi energy, a strong filling-factor dependence of the substrate-induced Dirac mass, and also the smallest particle-hole asymmetry reported to date in graphene cyclotron resonance. 
\end{abstract}

\maketitle

In graphene, Coulomb interactions combine with spin and valley degrees of freedom to generate an approximate SU(4) symmetry, which when broken can give rise to novel magnetic ground states in the quantum Hall regime at high magnetic fields. These phenomena have been explored by a variety of experimental probes including electronic transport, quantum capacitance, and scanning probe microscopy experiments \cite{zhang_landau-level_2006,checkelsky_zero-energy_2008,song_high-resolution_2010,young_spin_2012,hunt_massive_2013,zibrov_even-denominator_2018,li_scanning_2019}. However, the excited states of this system due to collective excitations between Landau levels (LLs) in the broken symmetry regime have been little explored to date \cite{jiang_valley_2019,onodera_cyclotron_2020}. Graphene is an ideal platform in which to pursue such studies because, in contrast to traditional two-dimensional electron systems having a parabolic dispersion, the linear dispersion of graphene allows the contribution of many-particle interactions to directly modify the LL transition energies in measurements of the cyclotron resonance (CR). Thus the interplay of interaction effects and broken symmetries can be explored spectroscopically and on an even footing. 

In this work we study the CR in high quality monolayer graphene by varying the LL filling factor at various fixed values of the magnetic field. Several interband transitions are observed to display non-trivial dependences on the filling factor. In the lowest interband transition, an intriguing pattern of resonances appears starting with a single peak at half-filling of the $n=0$ LL (ZLL), that then splits into four peaks at $\nicefrac{3}{4}$-filling, and reduces to just two as the level becomes completely occupied; meanwhile the higher interband excitations show interesting sequences of spectral weight shifts with the changing LL occupation. Using a simple model of transitions between LLs having interaction-enhanced valley and Zeeman gaps, we find the gap in the ZLL arising from coupling of graphene to the encompassing hexagonal boron nitride (hBN) becomes strongly enhanced both at half-filling and as the magnetic field is increased. Moreover, we observe an enhancement of Zeeman gaps both at and well below the Fermi level, with the latter indicating an indirect exchange coupling due to lattice-scale interactions coupling the two valleys in graphene. While this work specifically addresses physics in graphene, the approach is applicable in principle to any system with a linear dispersion and so may find utility in understanding the competing roles of interactions and symmetry breaking in Dirac, Weyl, or strongly correlated materials \cite{rao_cyclotron_2019-1}.

In a strong magnetic field and absent symmetry breaking, graphene develops four-fold degenerate LLs (two each for electron spin and the $K$ and $K'$ valleys) with single-particle energies given by $E_n=\textrm{s}_n \hbar \omega_c \sqrt{|n|}$, where $\omega_c=\sqrt{2} v l_B^{-1}$ is the cyclotron frequency, $v{\sim}10^6$ m/s is the band velocity, $l_B=\sqrt{\hbar/e B}$ the magnetic length, $\textrm{s}_n=\textrm{sign}(n)$, and $n=0,\pm1,\pm2...$ is the orbital index \cite{mcclure_diamagnetism_1956,gusynin_anomalous_2007}. If the sublattice symmetry of graphene is broken, as is common for hBN-encapsulated devices, the valley-polarized $n=0$ level is split by $E_{0,K}(E_{0,K'})=+(-)M$, where $M$ is the Dirac mass  \cite{gusynin_unusual_2006}, and the $|n|>0$ levels are shifted according to $E_n=\textrm{s}_n \hbar \omega_c \sqrt{|n|+\mu^2}$, with $\mu=M/\hbar\omega_c$. The CR energies of inter- or intra-band transitions from LL $m$ to $n$ are then given by the level separation 
\begin{equation}
\Delta E_{m,n}=\hbar \omega_c \ \left(\textrm{s}_n \sqrt{|n|+\mu^2}\ - \textrm{s}_m \sqrt{|m| + \mu^2}\ \right). \label{eq1}
\end{equation}
\noindent with the selection rule $|n|-|m|=\pm1$. In graphene, these energies can also include contributions from many-particle interactions, in contrast to materials with a parabolic dispersion where the center-of-mass and inter-particle coordinates are separable and CR becomes insensitive to electron interactions, a result known as Kohn's theorem \cite{kohn_cyclotron_1961, kallin_many-body_1985,throckmorton_failure_2018}. The linear dispersion of graphene mixes these coordinates so that interactions can directly impact LL transitions \cite{iyengar_excitations_2007,bychkov_magnetoplasmon_2008,shizuya_many-body_2010,roldan_spin-flip_2010,faugeras_landau_2015,sonntag_impact_2018}, leading to deviations from Eq.\ \ref{eq1} \cite{jiang_infrared_2007,henriksen_interaction-induced_2010,chen_observation_2014,jiang_valley_2019,nedoliuk_colossal_2019} and a dependence of CR on the LL filling factor, $\nu=2 \pi n_s l_B^2$, where $n_s$ is the charge carrier sheet density \cite{russell_many-particle_2018}.

The sample used in this study is an 820 $\mu$m$^2$ sheet of monolayer graphene sandwiched between ${\approx}40$-nm-thick flakes of hexagonal boron nitride, assembled using a dry-stacking technique \cite{wang_one-dimensional_2013} and placed on a 4-nm-thick flake of single-crystal graphite lying on a lightly-doped, oxidized Si wafer. Electrical contacts to the edge of the graphene were made using 3/60-nm-thick films of Cr/Au, defined by standard electron beam lithography fabrication. A 90-$\mu$m aluminum foil aperture restricts the infrared light to the region immediately surrounding the sample. All spectroscopic data in this work were acquired at a base temperature of 300 mK (estimated sample temperature of $<2$ K \cite{Note1}) for fixed values of the magnetic field using a broadband Fourier-transform infrared spectrometer with instrumental resolution of 0.5 meV (with exploratory traces at other resolutions~\cite{Note1}). Unpolarized blackbody light from the spectrometer was coupled through a KBr window into a cryogen free dilution refrigerator with a 14 T solenoid, focused to and defocused from the sample using custom parabolic optics, and funneled via a compound parabolic collector to a composite Si bolometer. Traces are acquired at target LL filling factors and normalized to spectra taken at much higher $\nu$ where many of the transitions at the target $\nu$ are Pauli-blocked, so that absorption features common to both traces divide to unity \cite{Note1}. Each normalized spectrum was averaged for approximately four hours. 

\begin{figure*}[t]
\includegraphics[width=\textwidth]{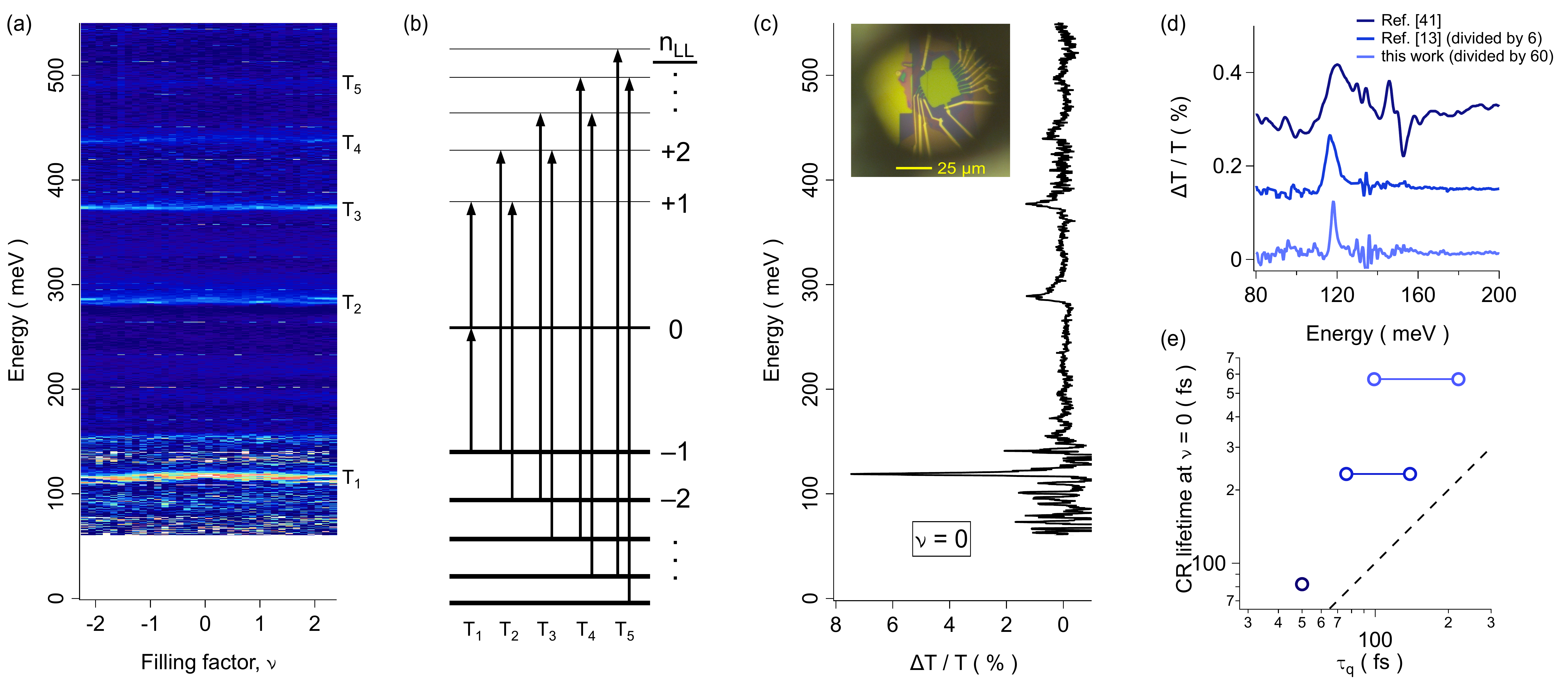}
\caption{\textbf{Cyclotron resonance transitions in graphite-gated monolayer graphene at $B=8$ T.} (a) Color map of the normalized transmission spectra $\Delta T/T$ in the mid-infrared as a function of the LL filling factor, $\nu$, measured in the device shown inset to (c). Several sharp CR transitions are visible, labeled $T_1$ through $T_5$. The higher noise in the region of $T_1$ is due to overall lower transmission in the 60-150 meV energy range, compared to the other transitions (and below 60 meV the signal goes to zero) \cite{Note1}. (b) Schematic showing the allowed Landau level transitions at $\nu=0$, consisting of nominally degenerate pairs. (c) Representative linecut of the color map at $\nu=0$. (d) Evolution of CR lineshape at $\nu=0$ with sample quality: the top trace is from a graphene-on-SiO$_2$ device with mobility of $17,000$ cm$^2$/Vs \cite{henriksen_interaction-induced_2010}, the middle is from an hBN-encapsulated device on SiO$_2$ \cite{russell_many-particle_2018}, and the bottom is the present graphite-gated, hBN-encapsulated device; the latter two have the same mobility, $\mu \approx 200,000$ cm$^2$/Vs. (e) CR lifetime $\tau_{CR}=\hbar/\Gamma$ at $\nu=0$ ($\Gamma$ the half-width at half-max) vs a spread of quantum scattering times $\tau_q$, derived from Shubnikov-de Haas oscillations acquired for a range of carrier densities at 3 K (colors correspond to traces (d)) \cite{Note1}. The dashed line marks $\tau_{CR}=\tau_q$.   \label{f1}} 
\end{figure*}

In graphene, several interband CR transitions $T_i$ can be observed simultaneously at fixed filling factor, comprising nominally degenerate pairs of inter-LL excitations $n=-i\to i-1$ and $1-i\to i$ with energies given by Eq.\ \ref{eq1}. Figure \ref{f1}(a) shows a color map of transitions $T_1$ through $T_5$ acquired as a function of $\nu$, in which the square-root dependence of the energies on the LL indices is immediately apparent. A schematic of the allowed transitions at half-filling is drawn in Fig.\ \ref{f1}(b), and a representative linecut at $\nu=0$ is shown in Fig.\ \ref{f1}(c). The very narrow resonances follow from recent improvements in sample fabrication \cite{dean_boron_2010,zibrov_tunable_2017}, and are key to enabling our observations. In Fig.\ \ref{f1}(d) we show $T_1$ at $\nu=0$ in devices from the present and two prior works \cite{henriksen_interaction-induced_2010,russell_many-particle_2018}, revealing a clear decrease in the half-width at half-max, $\Gamma$. In fact, the lower two traces in Fig.\ \ref{f1} (d) provide a comparison of two common gating methods: the middle trace is acquired in a sample with a distant, doped Si/SiO$_2$ substrate on which the encapsulated monolayer rests \cite{russell_many-particle_2018}, while the lower trace from the present work uses a local graphite gate. By chance these two devices have similar charge carrier mobilities of 200,000 cm$^2$/Vs, but the graphite-gated device shows greater values of the quantum scattering time $\tau_q$ extracted from Shubnikov-de Haas oscillations \cite{coleridge_low-field_1989}. This likely reflects improved screening of charged impurities in the SiO$_2$ by the graphite. The CR lifetimes $\tau_{CR}=\hbar/\Gamma$ in Fig.\ \ref{f1} (e) are similarly improved, and in fact the value of $\sim$600 fs quoted for the present device is a lower limit as even narrower lines with $\tau_{CR}\approx 2.5$ ps ($\Gamma=0.26$ meV) are seen at higher instrumental resolution. This latter value is close to the transport time derived from the mobility \cite{Note1}, suggesting that impurity collisions limit the CR lifetime. Consistent with prior observations of CR in AlGaN/GaN heterostructures \cite{syed_electron_2004}, $\tau_{CR}$ can be several times larger than $\tau_q$ which is reduced by variations in the carrier density across the sample. 

\begin{figure*}[t]
\includegraphics[width=\textwidth]{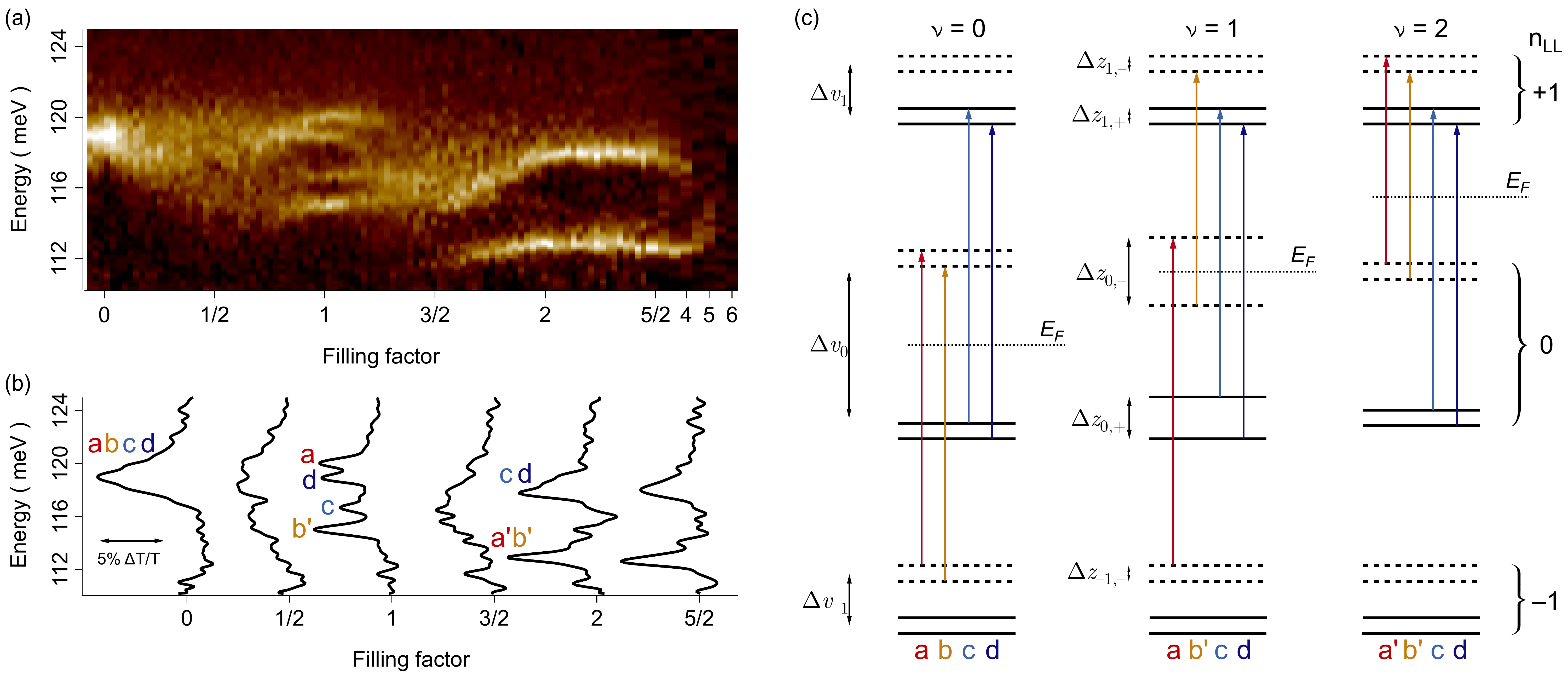}
\caption{\textbf{Evolution of transition $T_1$ vs filling factor.} (a) High-resolution map of $T_1$ vs filling factor from $\nu=-0.07$ to $\nu=+2.5$. Traces were acquired every $\delta \nu=0.026$, with additional traces at $\nu=3, 3.5,...6$. Starting with a single bright peak at $\nu=0$, four peaks appear near $\nu=+1$ which reduce to two peaks at $\nu=2$ and higher. By $\nu=6$ the $T_1$ transition is extinguished as the participating LLs are filled. (b) Detail of transitions by linecuts at half-integer fillings. The linewidths at $\nu=+1$ are the narrowest observed with $\tau_{CR}$ reaching 2.5 ps, or a resonance quality factor $Q=220$. In between integer values of $\nu$, only a single broad resonance is resolved. (c) Schematic of transitions involving the $n=-1, 0,$ and $+1$ LLs. Solid (dashed) lines indicate the $K$ ($K'$) valleys, with valley gaps $\Delta v_i$ and spin splittings $\Delta z_i$ explicitly included. The Fermi energy $E_F$ is shown as a dotted line. Each of the four spin and valley-preserving CR transitions are shown in different colors corresponding to the labels $a$, $b$, $c$, and $d$. As the filling factor is increased, the two transitions from $n=-1$ to $0$ become Pauli-blocked and are replaced by transitions from $n=0$ to $+1$; this is indicated by a label change $a,b\to a',b'$.  Gaps indicated in this schematic represent single particle levels enhanced by electron-electron interactions as discussed in the text. \label{f2}} 
\end{figure*}

In Figure \ref{f2} we focus on the $T_1$ transition over filling factors $\nu=0$ to $+6$, where a marked non-monotonic evolution is seen from a single resonance at $\nu=0$, to four resonances around $\nu=+1$, which reduce back to two for $\nu \gtrsim +2$ that both fade away as $\nu \to 6$ and the $n=+1$ LL is completely filled; a sudden sharp rise in the lower energy resonance above $\nu=5$ presages the extinction of the resonance. Linecuts in Fig.\ \ref{f2}(b) show details at half-integer $\nu$. The resonances manifest in intriguing patterns: the higher energy peaks at $\nu=1$ appear and disappear at different $\nu$ values, while the lower energy pair appear simultaneously and then merge with increasing $\nu$. At $\nu=2$ the upper peak first appears at a lower energy near $\nu=3/2$ and then rapidly rises before leveling off for $\nu\gtrsim2$. Note these features at $\nu=1$ and 2 persist over a wide range of $\nu$. This is a real effect and not due, for instance, to small variations in the carrier density across the sample: from the width of the Dirac peak in the zero-field resistance vs density, we estimate a distribution of carrier densities $\delta n_s \approx 2\times 10^{10}$ cm$^{-2}$, or $\delta \nu \approx0.1$ at 8 T, rather smaller than the range over which the $\nu=1$ and $2$ features persist \cite{Note1}. At $\nu=1/2$ and $3/2$, broad resonances appear that nevertheless maintain the full spectral weight, suggesting all transitions are present but undifferentiated \cite{Note1}. This could indicate the presence of dark magnetoexciton modes serving as additional scattering channels: there are up to 16 distinct transitions between the $0$ and $\pm1$ LLs  although only the four that conserve spin and valley are optically active \cite{iyengar_excitations_2007}.

In Figure \ref{f2}(c) we introduce schematics representing the simplest model of transitions between the $n=0$ and $\pm1$ LLs that aligns with the observed CR. These are drawn for $\nu=0,+1$, and $+2$, with each of the four spin- and valley levels shown explicitly albeit with greatly exaggerated level shifts and gap sizes. In graphene, the inapplicability of Kohn's theorem implies the CR transition energies will reflect the single-particle LL separations plus many-particle shifts of the levels, along with excitonic and exchange corrections due to the excited electron and remnant hole \cite{iyengar_excitations_2007,bychkov_magnetoplasmon_2008,shizuya_many-body_2010,shizuya_many-body_2018,sokolik_many-body_2019}. Of course the measured energies do not indicate which portion is due to level shifts vs exciton corrections. Therefore we model each transition energy as a sum of the LL separation plus the difference of any valley and Zeeman gaps in each level, with the understanding that these gaps are meant to represent both single- and many-particle energies.

For instance, at $\nu=+2$ two resonances are observed although up to four transitions (two each for valley and spin, labeled $a'$, $b'$, $c$, and $d$ in Fig.\ \ref{f2}(c)) are allowed. All LLs are either completely filled or empty so that interactions are expected to be minimized. If we assume that the Zeeman splittings in the $n=0$ and $1$ LLs are equal, then the observed CR splitting $\Delta E^{\nu=2} = E^{c,d} - E^{a',b'}$ arises from transitions originating on either side of the valley gap in $n=0$. Note if the $n=1$ LL also has a non-zero valley gap, it is still the \emph{difference} of these gaps,  $\Delta v^{\nu=2} \equiv \Delta v_0 - \Delta v_1=\Delta E^{\nu=2}$ that is detected. If the Zeeman splitting were also enhanced in one level over another, this picture would predict additional resonances not present in the data. Fitting the two peaks at $\nu=2$ with Lorentzians, we find $\Delta E^{\nu=2}= 5.0(1)$ meV. Since any valley splitting of the $n=1$ LL is likely to be small, this should be a good measure of the valley gap in the ZLL. We identify this gap as due to sublattice symmetry-breaking from the presence of hBN \cite{hunt_massive_2013}, and calculate a Dirac mass $M=2.5$ meV.

At $\nu=0$, the single peak indicates the four allowed transitions are all degenerate. By the schematic in Fig.\ \ref{f2}(c), the CR energy is given by the LL separation plus half the difference of the valley gaps in the zeroth and $\pm1$ LLs. That a single resonance is seen implies the valley gaps in the $n=\pm1$ levels must be equal, and all of the Zeeman gaps must also be the same, or else additional CR lines would be seen. Actually, the $\nu=0$ resonance is the broadest in $T_1$, suggesting there may be unresolved lines due either to a differential enhancement of these gaps, or a level repulsion between the two degenerate pairs labeled \{$a$,$b$\} and \{$c$,$d$\} in the figure if lattice-scale interactions couple the $K$ and $K'$ valleys. Indeed, such a splitting appears at 13 T as discussed below. For now we determine the valley gap difference to be $\Delta v^{\nu=0} = 2\ (E^{\nu=0} - E^{\nu=2}_{avg})=7.3(5)$ meV, where $E^{\nu=2}_{avg}$ is the average energy of the two peaks at $\nu=2$. This yields a Dirac mass of $3.7$ meV, substantially enhanced over its value at $\nu=2$.

Finally, four resonances are seen at $\nu=1$, which requires each transition to comprise a unique combination of valley and spin gaps in the initial and final LLs. In Fig.\ \ref{f2}(c) we sketch a scenario where, for instance, the two transitions \{$c,d$\} (that are degenerate at $\nu=0$ and $+2$), now gain distinct energies at $\nu=1$ when the Zeeman gaps in the $n=0$ and $1$ LLs become unequal. Moreover, the two Zeeman gaps in the $n=0$ level marked $\Delta z_{0,-}$ and $\Delta z_{0,+}$ must be differentially enhanced, or else the transitions marked $a$ and $d$ will remain degenerate. The difference of the valley gap energies in the $n=0$ and $\pm1$ LLs, namely $\Delta v^{\nu=1} = \Delta v_0 - \Delta v_{\pm1}$, and the two Zeeman differences $\Delta z_- = \Delta z_{0,-}-\Delta z_{\pm1,-}$ and $\Delta z_+ = \Delta z_{0,+}-\Delta z_{1,+}$, can be extracted by inverting a matrix that records the contribution of each gap to the transition energy. The full procedure is described in the Supplemental Material and yields $\Delta v^{\nu=1} = 5.0 ~\textrm{meV}, \Delta z_+ = 2.1~ \textrm{meV}$, and $\Delta z_- = 4.3 ~\textrm{meV}$. Note we assume that gaps in the $n=\pm1$ levels are identical. While the size of this valley gap is close to that found at $\nu=+2$, the spin gaps are significantly larger than the bare Zeeman energy at this field, $E_Z=0.93$ meV, indicating a clear role for electron interactions. The enhanced $\Delta z_+$ splitting is notable, as both levels are occupied and well below the Fermi energy. This is reminiscent of indirect exchange splitting in the spin sector seen in GaAs quantum wells \cite{dial_high-resolution_2007}, except that here the splittings occur in different valleys, indicating the presence of lattice-scale interactions coupling valleys $K$ and $K'$. Meanwhile the size of $\Delta z_-$ at the Fermi energy compares well to a transport gap of $\sim 5$ meV, for $\nu=-1$ at 9 T, found in Ref.\ \cite{young_spin_2012}. Casting these as effective $g$-factors, we find the spin gap at the Fermi level has $g_{z;-}^* = \Delta z_-/\mu_B B = 9.3$, and the buried spin gap (in the $K$ valley) has $g_{z;+}^* = 4.5$.

We briefly note that although Kohn's theorem does not hold in graphene in general, a limited version is predicted to survive for the $T_1$ transition \cite{iyengar_excitations_2007,bychkov_magnetoplasmon_2008,shizuya_many-body_2010}. However, the filling-factor-dependent shifts and splittings found here strongly imply that even this remnant does not hold. We speculate that either the hBN-induced moir\'e pattern (with a length scale comparable to the magnetic length), or the lattice-scale interactions invoked to explain the $\nu=0$ ground state \cite{zibrov_even-denominator_2018}, are sufficient to break translation invariance and render Kohn's theorem inoperable.

\begin{figure*}[t]
\includegraphics[width=\textwidth]{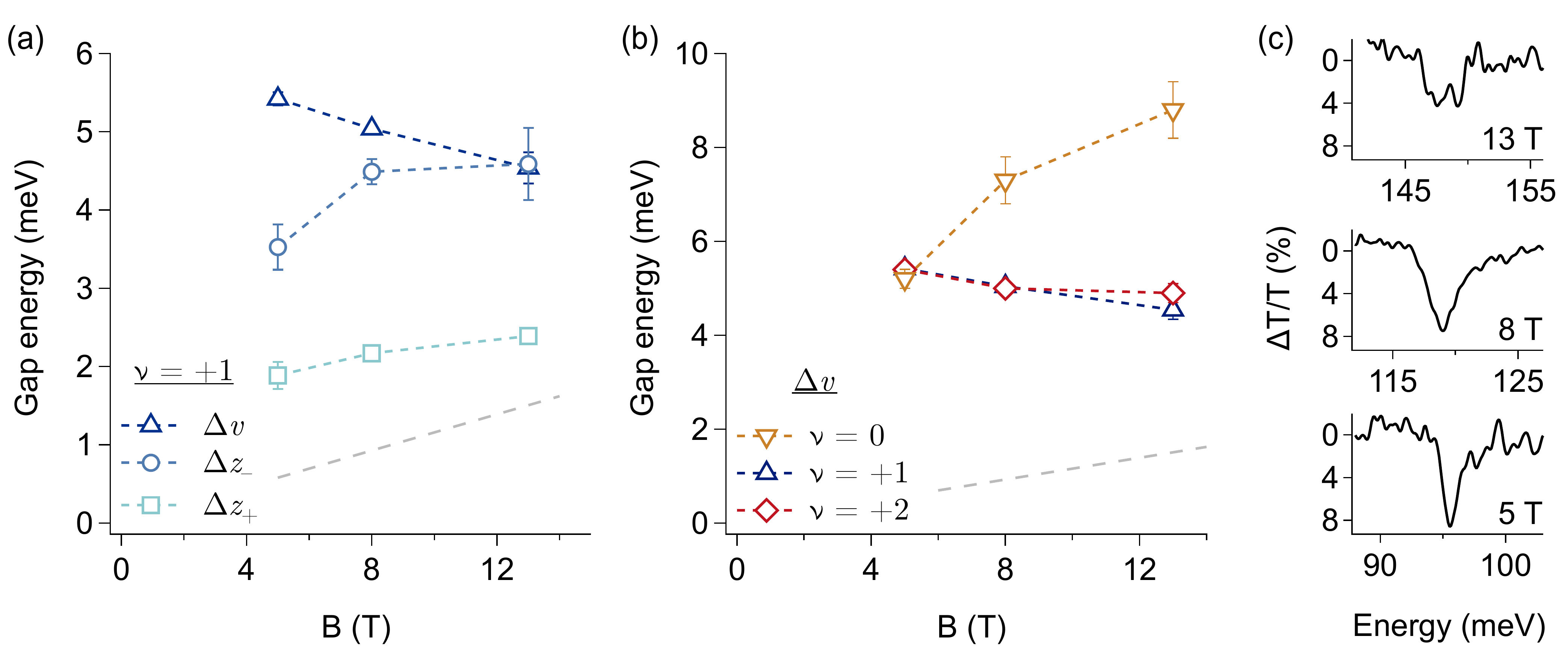}
\caption{\textbf{Evolution of splittings with magnetic field.} (a) The magnetic field dependence of the observed gaps at $\nu=1$.  The valley gap $\Delta v$ appears to decrease with increasing field while the spin gaps $\Delta z_{-}$ and $\Delta z_{+}$ increase with increasing field.  All splittings observed are significantly larger than the bare Zeeman energy shown as a gray dashed line. (b) Comparison of the valley gaps calculated at integer filling as a function of magnetic field. (c) The $T_1$ resonance at $\nu=0$ for three magnetic fields.  With increasing field, the resonance broadens and shows an incipient splitting.  \label{f3}} 
\end{figure*}

To better understand the nature of these splittings, we show the magnetic field dependence of the extracted spin and valley gaps at $\nu=1$ in Fig.\ \ref{f3} (a). The measured spin gap energies are substantially larger than the Zeeman energy, which suggests an interaction enhancement consistent with the ferromagnetic ground state at quarter filling in Ref.\ \cite{young_spin_2012}. The gaps exhibit a sub-linear increase with magnetic field, close to the $\sqrt{B}$ dependence expected for interaction effects, although further work is needed to understand the precise field dependence. Unlike the spin gaps, the $\nu=1$ valley gap is observed to decrease with increasing magnetic field. In Fig.\ \ref{f3} (b) this valley gap is compared with those for the half- and fully-filled $n=0$ LL, where we find the gaps at $\nu=1$ and $2$ remain closely matched as the field changes. Since interaction effects should be weakest at $\nu=2$, this agreement suggests the valley gap at $\nu=1$ is hardly impacted by interactions. In contrast, the valley gap extracted for $\nu=0$ increases dramatically with increasing magnetic field, consistent with the understanding that the ground state at $\nu=0$ involves an interaction-driven breaking of valley symmetry which drives an enhancement of the gap \cite{kharitonov_phase_2012,young_spin_2012,zibrov_even-denominator_2018}. A closer look at $T_1$($\nu=0$) for multiple fields in Fig.\ \ref{f3} (c) shows the resonance broadens at 8 T compared to 5 T, and develops a clear splitting by 13 T (note $\Delta v^{\nu=0}$ in Fig.\ \ref{f3} (b) uses the average value of this splitting). As noted previously, this is perhaps due to level repulsion of degenerate transitions in the two valleys by short-ranged Coulomb interactions, known to be important in the study of quantum Hall ferromagnetism but not yet studied in the context of CR in graphene \cite{kharitonov_phase_2012}.

\begin{figure*}[t]
\includegraphics[width=0.5\textwidth]{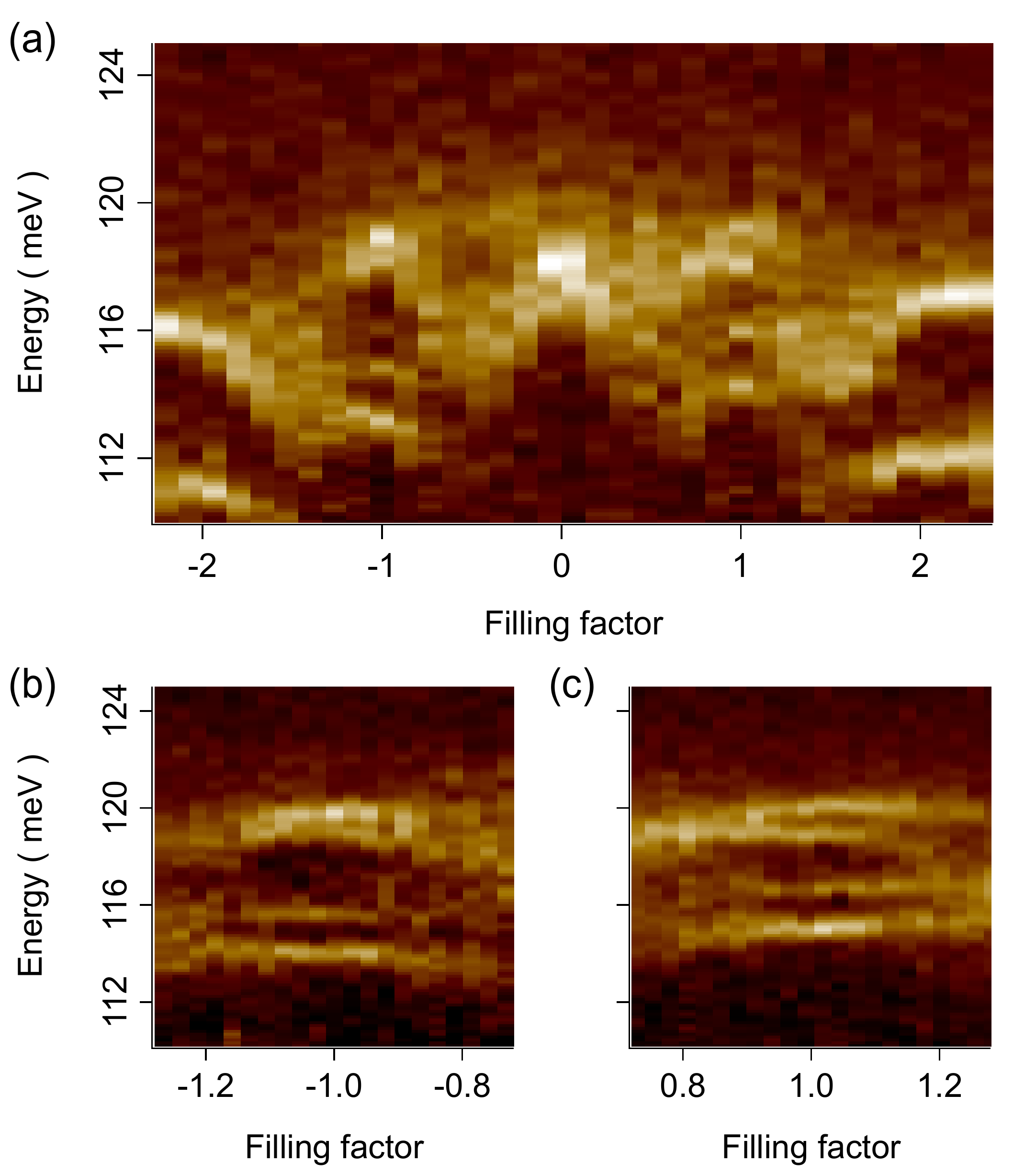}
\caption{\textbf{Particle-hole asymmetry.} (a) Closeup of $T_1$ transition from Fig.\ \ref{f1}(a), showing a clear asymmetry  for positive vs.\ negative filling factors. (b) and (c) show higher resolution (finer abscissa spacing) maps at $\nu=-1$ and $+1$, respectively, revealing particle-hole symmetry breaking in the four-fold splittings.  \label{f4}} 
\end{figure*}

In Fig.\ \ref{f4}(a) we zoom out to show $T_1$ over an equal range of positive and negative filling factors, and find a small but clear particle-hole asymmetry. For example, while the size of the $\nu=\pm2$ splittings are virtually identical at 5.0 meV, the hole-side peaks lie a full 1.0 meV lower in energy. Moreover a closer look at $\nu=-1$ and $+1$ in Fig.\ \ref{f4}(b) and (c) shows the two lower energy transitions are both separated by 1.7 meV and exhibit a slow ramp up with increasing $|\nu|$, but the hole-side pair is found 1.1 meV lower than the electron-side pair. Meanwhile, the two higher-energy peaks on the hole side nearly overlap, compared to $\nu=+1$ where we have seen they are individually resolved. Additionally, the relative shift of these higher-energy peaks with increasing $|\nu|$ shows opposing trends near $\nu=-1$ and $+1$, with both pairs lying close together at the left side of the graphs (more negative $\nu$) and separating toward the right (for more positive $\nu$), breaking $\nu \to-\nu$ symmetry. Finally, the highest energy peak on the hole side is only 0.4 meV lower than the electron side. Relative to the CR energy, this symmetry breaking is a $\sim0.8\%$ effect, too small to have been noticed in early broadband spectroscopic studies \cite{jiang_infrared_2007} but matching an asymmetry apparent in the data of Ref.\ \cite{henriksen_interaction-induced_2010,russell_many-particle_2018}.  However in terms of the many-particle-enhanced valley and spin gaps, these small shifts are quite significant. For instance, applying the same analysis used in the discussion of Fig.\ \ref{f2}, we find for $B=8$ T that $\Delta v^{\nu=-1} = 5.7$ meV;  $\Delta z_+^{\nu=-1} = 1.2$ meV (or effective g-factor $g^*_{z;+}=2.6)$; and $\Delta z_-^{\nu=-1}=6.7$ meV ($g^*_{z,-}=14.4$).

Such particle-hole asymmetry is not predicted by many-particle theories to date, but may arise at the single-particle level due to next-nearest-neighbor hopping \cite{peres_electronic_2006}. In this picture, a field-dependent asymmetry between the $-n\to n-1$ and $1-n\to n$ transitions was derived for the high-$n$ limit in Ref.\ \cite{plochocka_high-energy_2008}, giving $E_{asym}=3 \sqrt{2}\hbar \omega_c t' a/t l_B \approx 0.56$ meV at 8 T (where $t\ (t')$ is the nearest (next-nearest) neighbor hopping, and $a$ the C-C atom distance in graphene). This value lies within a factor of two of the asymmetry energies seen here, suggesting we are seeing an intrinsic property of the underlying band structure. In contrast, far larger particle-hole asymmetries up to a few percent of $E_{0,1}$ have been reported in swept-field CR studies of graphene-on-oxide, monolayer and multilayer epitaxial graphene, and encapsulated graphene with double moir\'e potentials \cite{deacon_cyclotron_2007,jiang_valley_2019,onodera_cyclotron_2020,nakamura_quantum_2020}.

\begin{figure*}[t]
\includegraphics[width=0.8\textwidth]{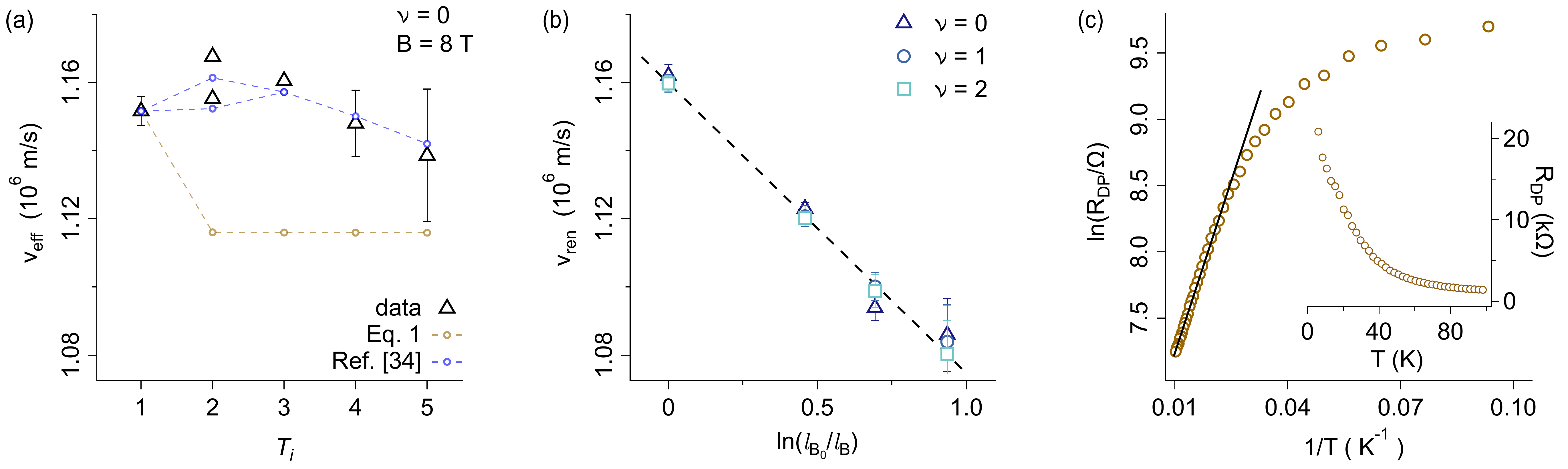}
\caption{\textbf{Effective velocity, renormalized velocity, and activated transport.} (a) The transition energies parameterized as effective velocities $v_{eff}=\Delta E^{meas}(T_i)/\Delta E^{calc}(T_i)[v=10^6;\mu=0]$ at $\nu=0$. The dashed lines predict $v_{eff}$ by two approaches: the tan lines show the single-particle model of Eq.\ \ref{eq1} including a finite Dirac mass, while the blue lines are calculated using the many-particle theory of Ref.\ \cite{shizuya_many-body_2018}. (b) The renormalized Fermi velocity $v^{ren}$, extracted from fits to the theory of Ref.\ \cite{shizuya_many-body_2018}, is plotted vs the logarithm of $\sqrt{B}$. The resulting linear dependence is the magnetic field equivalent of the zero-field velocity renormalization in graphene \cite{gonzalez_non-fermi_1994,shizuya_many-body_2010,elias_dirac_2011,faugeras_landau_2015}.(c) Arrhenius plot of the device resistance at charge neutrality and zero magnetic field; the slope implies a gap of 15.0 meV. Inset shows the measured resistance vs temperature.  \label{f5}} 
\end{figure*}

In Figure \ref{f5}(a) the transition energies at $\nu=0$ and 8 T are plotted as a function of transition number $T_i$, parameterized as an effective velocity $v_{eff}(T_i)=\Delta E^{meas}(T_i) / \Delta E^{calc}(T_i) [v=10^6;\mu=0]$. We see that $v_{eff}$ rises from $T_1$ to $T_2$ and thereafter gradually decreases in agreement with previous measurements \cite{russell_many-particle_2018}. We fit these data in two ways: first, in the basic non-interacting picture with energies given by Eq.\ \ref{eq1}, using fixed band velocity $v$ and mass $\mu$ (with $\mu$ set to the splitting at $\nu=2$). This model clearly does not capture the measured variation in $v_{eff}$. Far better results are found using the theory of Ref.\ \cite{shizuya_many-body_2018} which accounts for many-particle contributions to CR in a single-mode approximation \cite{macdonald_magnetoplasmon_1985}. The fit has three parameters: an interaction-renormalized band velocity $v^{ren}$, the Dirac mass, and an overall Coulomb interaction we fix at $V_C=\sqrt{\pi/2}\ e^2/(4 \pi \epsilon l_B)=50$ meV \cite{Note2}. This provides a good account of the variation in $v_{eff}$ vs $T_i$ and also the size of the $T_2$ splittings, and yields $v_{ren}=1.105\times10^6$ m/s and $M=2.76$ meV, close to the value at $\nu=2$. Carrying out this procedure at other magnetic fields and filling factors yields the $v^{ren}$ values in Fig.\ \ref{f5}(b). There, the resulting linear decrease against $\text{ln}(\sqrt{B/B_0})$ is anticipated in Ref.\ \cite{shizuya_many-body_2010}, which predicts the slope is given by $-(\alpha c/4\epsilon)$ where $\alpha$ is the fine structure constant and $c$ the speed of light. This running of the velocity with $B$ is the generalization to finite field of the interaction-renormalized band velocity predicted before graphene was isolated \cite{gonzalez_non-fermi_1994} and seen in electronic transport \cite{elias_dirac_2011}. The slope determines a dielectric constant of $\epsilon = 6.4$, which is likely dominated by the in-plane $\epsilon$ of hexagonal boron nitride \cite{laturia_dielectric_2018}, and is in good agreement with magneto-Raman measurements \cite{faugeras_landau_2015}.

\begin{figure*}[t]
\includegraphics[width=0.7\textwidth]{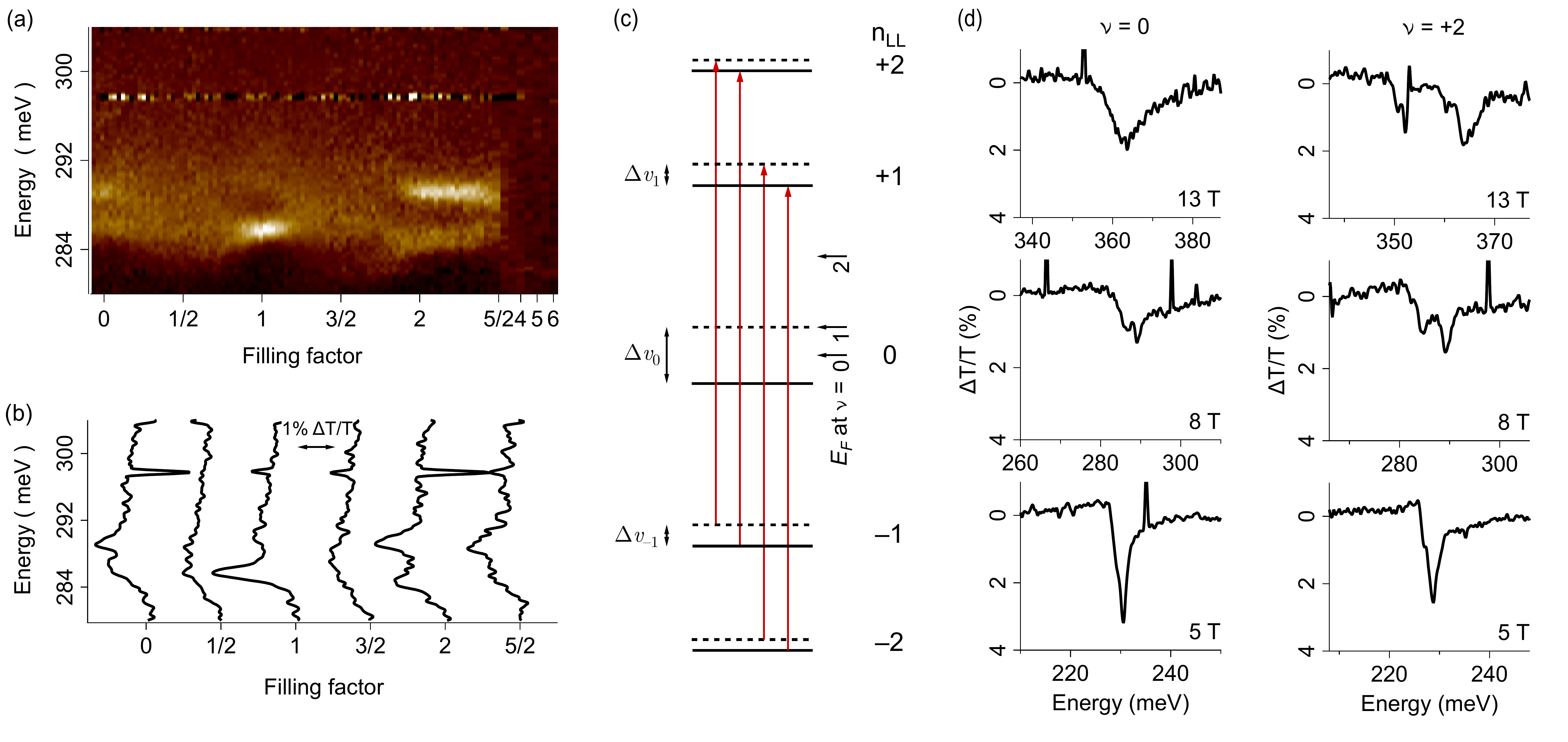}
\caption{\textbf{Evolution of the second interband transition $T_2$ vs filling factor.} (a) High-resolution map of $T_2$ for the same filling factor range as Fig.\ \ref{f1}. The horizontal band at 298 meV is due to harmonics of 60 Hz. Two peaks are just resolved at $\nu=0$, while the spectrum is dominated by a single peak at $\nu=+1$, and two peaks appear again at $\nu=+2$ albeit with more intensity in the higher peak. The resonances are broader than those at $T_1$, and exhibit a high energy tail which may be a result of multiple reflections in the substrate \cite{henriksen_interaction-induced_2010}. (b) Linecuts at half-integer filling factors. (c) Schematic of $T_2$ transitions, with the $K$ ($K'$) valley shows as a solid (dashed) line. Zeeman splittings are suppressed. (d) Spectra at $\nu=0$ and $+2$ as a function of magnetic field. A remarkably large splitting, nearly 13 meV in size, appears at $\nu=+2$ at 13 T. \label{f6}} 
\end{figure*}

Finally, in Figure \ref{f6} we explore the evolution of the second interband transition, $T_2$. Inspection of the color map and linecuts shows that a splitting is just resolved at $\nu=0$, with peaks of approximately equal strength. This evolves into a bright and sharp peak at $\nu=+1$ accompanied by a much weaker resonance on the high energy side, while at $\nu=+2$ the splitting persists but most of the spectral weight has shifted to the higher energy peak. Similar to $T_1$, at half-integer fillings only a single broad resonance is seen although the integrated intensity remains constant over this range of $\nu$ \cite{Note1}. The peaks are split by 2.8 meV at $\nu=0$ and 4.7 meV at $\nu=+2$. The behavior with changing magnetic field shown in Fig.\ \ref{f6} (d) is rather different than for $T_1$. For $T_2$, a single $\nu=0$ peak at 5 T gains a splitting at 8 T but reverts to a broader single resonance at 13 T. Since the $T_2$ transition comprises two nominally degenerate pairs of transitions $n=-2\to+1$ and $-1\to+2$ in each valley, as above a weak valley coupling may split the degeneracy. Whether the splitting is observed may depend on the width of the resonances, which increases with field. For instance at 8 T, the splitting is greater than the width and can be seen, but is likely masked by further broadening of the resonance by 13 T. In contrast, at $\nu=2$, the sharp single peak at 5 T evolves by 13 T into an unexpectedly large splitting, nearly 13 meV, far larger than any other splitting seen in this work. In the many-particle theories of Ref.\ \cite{bychkov_magnetoplasmon_2008,shizuya_many-body_2018}, interactions alone suffice to break the degeneracy of the $n=-2\to+1$ and $-1\to+2$ transitions at both $\nu=0$ and $2$; further small corrections are expected for a finite Dirac mass. For $T_2$, Ref.\ \cite{shizuya_many-body_2018} predicts an approximately 3 meV splitting for a 5 meV gap. This roughly matches the scale of splittings at 8 T, but greatly underestimates the $\nu=2$ splitting at 13 T. This large splitting is a surprise, since for $\nu=2$ all orbital levels are filled or empty and interaction corrections should be minimal. At this time no mechanism is clearly responsible for such a large splitting at $\nu=2$.

The next higher interband transition, $T_3$, also shows an intriguing and larger-than-expected sequence of splittings. A map of the transition energies vs filling factor along with linecuts at integer filling factor is included in the Supplementary Material. The signal-to-noise in even higher interband transitions is not sufficient to resolve splittings.

When applied to graphene, cyclotron resonance becomes a novel tool for spectroscopy of many-particle physics since Kohn's theorem no longer applies. Here it enables us to follow the evolution of many-particle enhanced gaps in the broken symmetry regime of clean monolayer graphene, where we find a Dirac mass that is significantly enhanced at half-filling of the zeroth LL, and Zeeman gaps both at and below the Fermi energy that are enhanced by direct or indirect exchange effects. These observations highlight the importance of lattice-scale interactions coupling the $K$ and $K'$ valleys in graphene. Moreover, a very small but finite particle-hole asymmetry is seen, that underscores the device quality and sets upper limits on the symmetry of the linear dispersion in graphene. These results promise that with continually improving device fabrication techniques, it will soon be possible to perform spectroscopy of excited states in the fractional quantum Hall regime.

\medskip

\noindent \textbf{Acknowledgements}
We acknowledge informative discussions with H.\ Fertig, A.\ MacDonald, and K.\ Shizuya, and are grateful for support from the Institute of Materials Science and Engineering at Washington University in St.\ Louis. KW and TT acknowledge support from the Elemental Strategy Initiative conducted by the MEXT, Japan, Grant Number JPMXP0112101001, JSPS KAKENHI Grant Number JP20H00354 and the CREST(JPMJCR15F3) JST.

\newpage

\clearpage
\clearpage

\newpage

\clearpage
\clearpage

\newpage

\clearpage
\clearpage

\newpage

\clearpage
\clearpage

\newpage

\clearpage
\clearpage

\newpage

\clearpage
\clearpage

\newpage


\bibliographystyle{nsf_erik}

\end{document}